\documentclass{acm_sen_article}
\usepackage{flushend}
\title{Survey Results on Threats To External Validity, Generalizability Concerns, Data Sharing and University-Industry Collaboration in Mining Software Repository (MSR) Research}
\numberofauthors{4}
\author{
\alignauthor Ashish Sureka\\
       \affaddr{Software Analytics Research Lab (SARL), India}\\
       \email{ashish@iiitd.ac.in}
\alignauthor Ambika Tripathi\\
       \affaddr{IIIT Delhi, India}\\
       \email{ambika1333@iiitd.ac.in}
\alignauthor Savita Dabral\\
       \affaddr{IIIT Delhi, India}\\       \email{savita1378@iiitd.ac.in}
}
\begin{document}

\maketitle
\begin{abstract}
Mining Software Repositories (MSR) is an applied and practise-oriented field aimed at solving real problems encountered by practitioners and bringing value to Industry. Replication of results and findings, generalizability and external validity, University-Industry collaboration, data sharing and creation dataset repositories are important issues in MSR research. Research consisting of bibliometric analysis of MSR paper shows lack of University-Industry collaboration, deficiency of studies on closed or propriety source dataset and lack of data as well as tool sharing by researchers. We conduct a survey of authors of past three years of MSR conference ($2012$, $2013$ and $2014$) to collect data on their views and suggestions to address the stated concerns. We asked $20$ questions from more than $100$ authors and received a response from $39$ authors. Our results shows that about one-third of the respondents always make their dataset publicly available and about one-third believe that data sharing should be a mandatory condition for publication in MSR conferences. Our survey reveals that more than $50\%$ authors used solely open-source software (OSS) dataset for their research. More than $50\%$ of the respondents mentioned that difficulty in sharing Industrial dataset outside the company is one of the major impediments in University-Industry collaboration.   
\end{abstract}

\keywords{Empirical Software Engineering, Mining Software Repositories (MSR), Software Engineering Dataset Repositories, Threats to External Validity, University-Industry Collaboration}

\section{Research Motivation and Aim}
Mining Software Repositories (MSR) is one of the fastest growing field and community within Software Engineering and consists of analysing the rich data available in software repositories to uncover interesting and actionable information about software systems and projects \cite{Lanza2012}\cite{Zimmermann2013}\cite{Devanbu2014}. MSR is data-driven and falls under Empirical Software Engineering (ESE). MSR is an applied and practise-oriented field aimed at solving real problems encountered by practitioners and bringing value to Industry. Due to the nature of the discipline and its objectives, there are several factors such as reproducibility or replication of findings or results, generalizability of approach to other dataset, data sharing by researchers and University-Industry collaboration which are crucial in MSR research. Tripathi et al. conduct a bibliometric analysis of past five years of research papers published in MSR series of conferences ($2010$-$2014$) and show that out of $187$ studies over a period of $5$ years, $90.9\%$ studies are conducted solely on OSS dataset \cite{tripathi2015}. Their findings indicate that only $14.43\%$ of the studies involve a University-Industry collaboration \cite{tripathi2015}.

The study presented in this paper is motivated by the need to gain a deeper understanding of the stated issues and their solution by conducting a survey of authors who have published papers in MSR conference. We conduct a survey consisting of $20$ questions of authors who have published papers in MSR $2012$, $2013$ and $2014$. We limit the scope of our analysis to only MSR series of conferences over the last three years. MSR research papers are also published in several other Software Engineering conference. However, selection of conferences and identifying MSR papers in such conferences by the authors can result in a selection bias. We eliminate selection bias by analysing publications only from MSR conference. While there have been bibliometric studies on MSR papers \cite{robles2010}\cite{tripathi2015} on the topic of replication, data sharing and University-Industry collaboration, the work presented in this paper is the first study involving a survey of MSR authors.

\section{Survey Questionnaire and Findings}
\begin{figure}[t]
\centering
\includegraphics[width=0.9\linewidth]{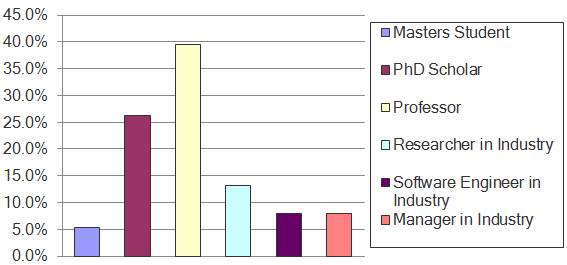}
\caption{Distribution of Survey Respondents across Job Roles}
\label{profile}
\end{figure}

We sent a survey consisting of $20$ questions to all authors of MSR $2012$, $2013$ and $2014$ conference and received a total of $39$ responses. We did not ask for their name or any personally identifiable information of the author. We have made the survey response publicly available as an Excel file\footnote{ http://bit.ly/1CXOV3r} so that other interested researchers can do analysis in addition to the findings presented in this paper. Table \ref{demographics}, Table \ref{msrexperience} and Figure \ref{profile} displays information on the work profile of the survey respondents. Table \ref{demographics} reveals that nearly $75\%$ of the survey respondents were affiliated to a University whereas only $25\%$ respondents were from Industry. Table \ref{msrexperience} and Figure \ref{profile} shows the distribution of the survey respondents across roles and job profiles. We received opinions from MS and PhD Scholars in University, Faculty Members, Researcher, Software Engineer and Manager in an Industry. While the percentage of Software Engineers and Managers [non-research roles] in Industry is small, there is still a representation. 
\begin{table}
\centering
\caption{Work Profile of Survey Respondents }
\label{demographics}
\begin{tabular}{|l|c|}
\hline
\multicolumn{2}{|p{7.5cm}|}{\textbf{[1] Are you currently working in an Industry or University?}} \\ \hline
Industry & 26.32\% \\
University & 73.68\% \\ \hline
\multicolumn{2}{|p{7.5cm}|}{\textbf{[2] What is your current job role within Industry or University?}} \\ \hline
Masters Student & 5.26\% \\
PhD Scholar & 26.32\% \\
Professor & 39.47\% \\
Researcher in Industry & 13.16\% \\
Software Engineer in Industry & 7.89\% \\
Manager in Industry & 7.89\% \\ \hline
\end{tabular}
\end{table}
\begin{table}
\centering
\caption{Work Experience in MSR and Authorship in MSR Conference}
\label{msrexperience}
\begin{tabular}{|l|c|}
\hline
\multicolumn{2}{|p{7.5cm}|}{\textbf{[3] How many years of experience do you have in Mining Software Repository (MSR) research?}} \\ \hline
0-2 years & 31.58\% \\
2-5 years & 28.95\% \\
More than 5 years & 39.47\% \\ \hline
\multicolumn{2}{|p{7.5cm}|}{\textbf{[4] How many publications (excluding data challenge track) you have in Mining Software Repository (MSR) series of conferences?}} \\ \hline
0-2 & 63.16\% \\
3-5 & 18.42\% \\
More than 6 & 18.42\% \\ \hline
\multicolumn{2}{|p{7.5cm}|}{\textbf{[5] How many Mining Software Repository (MSR) series of conferences have you attended?}} \\ \hline
0-2 & 65.79\% \\
3-5 & 18.42\% \\
More than 6 & 15.79\% \\ \hline
\end{tabular}
\end{table}
\subsection{University-Industry Collaboration}
University-Industry collaboration in Software Engineering (and particularly Empirical Software Engineering) is an area that has attracted several researcher's attention. There are both benefits and challenges associated with the collaboration. Runeson et al. present their experiences of a 10 year Industry-Academia collaboration program. Their study focuses on the time-horizon aspects of the Industry-Academia collaboration. Their study reveals that Industry time horizons are generally shorter compared to the academic perspective posing challenges to the collaboration \cite{Runeson2014}.  Martinez-Fernandez et al. present their practical experiences in designing and conducting empirical studies involving Industry-Academia collaboration. The focus of the collaboration described in their study is on Software Reference Architecture (SRA) projects in an IT consulting and services organization. Authors mention acquisition of realistic sources of data as well as creation of repeatable techniques and results as some of the major research challenges \cite{Martinez2014}. 

Enoiu et al. present an empirical exploration of enablers and impediments for collaborative research in Software Testing. They list open sharing of information for research purposes, use of a dedicated tooling platform and creation of a research culture as one of the major enablers. They mention resistance to change, lack of knowledge of techniques and tools evaluated in academia and not assuming stable research focus for the conduct of relevant experiments as the three main impediments \cite{Enoiu2014}. Runeson et al. present their experiences in a 2-year University-Industry collaboration project on software testing which involved on-site work by the researcher in the industry premises. They mention several factors which influence a successful collaboration: company management support, champion at the company, researcher's attitude and social skills and researcher's commitment to focus on industry needs \cite{Runeson2012}. Wohlin et al. presents a list of top 10 challenges (such as trust and respect, champion, social skills, commitment to company needs) to work with industry based on their experience from working with industry in a very close collaboration with continuous exchange of knowledge and information \cite{Wohlin2013}.
\begin{table*}
\centering
\caption{University-Industry Collaboration Success, Duration, Challenges and Suggestions}
\label{univindcoll}
\begin{tabular}{|l|c|}
\hline
\multicolumn{2}{|p{16cm}|}{\textbf{[6] Whether the University-Industry collaboration study was a success or a failure?}} \\ \hline
Failure & 12\% \\
Partially Successful & 72\% \\
Successful & 16\% \\ \hline
\multicolumn{2}{|p{16cm}|}{\textbf{[7] What is the average duration (in years) of your University-Industry Collaboration study?}} \\ \hline
0-1 years & 52\% \\
1-2 years & 44\% \\
More than 2 years & 4\% \\ \hline
\multicolumn{2}{|p{16cm}|}{\textbf{[8] What challenges you faced during the Collaboration?}} \\ \hline
Difficulty in sharing Industrial dataset outside the company & 52\% \\
Difference in focus on goal (business impact in Industry vs. scholarly impact in Academia) & 44\% \\
Different timeline (project deliverable timeline does not match academic milestones) & 4\% \\ \hline
\multicolumn{2}{|p{16cm}|}{\textbf{[9] Can you suggest some ways to improve the collaboration?}} \\ \hline
Industry sponsored PhD fellowships & 60\% \\
Encouraging student Internships & 64\% \\
Academic projects to be inclined towards Industry problems & 52\% \\
Others (please specify) & 12\% \\ \hline
\end{tabular}
\end{table*}
\begin{figure}[t]
\centering
\includegraphics[width=0.9\linewidth]{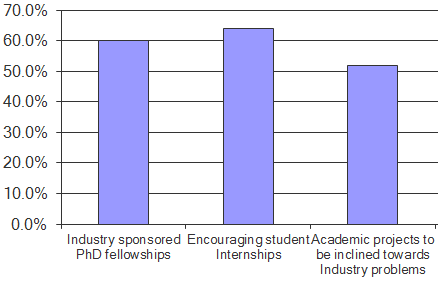}
\caption{Survey Respondents Opinion on Enabling University-Industry Collaboration}
\label{univind}
\end{figure}
\begin{figure}[t]
\centering
\includegraphics[width=0.9\linewidth]{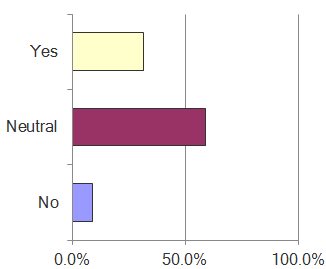}
\caption{Survey Respondents Opinion on Data Sharing as a Mandatory Condition for Publication}
\label{condition}
\end{figure}

We asked four questions related to University-Industry Collaboration to MSR authors. The questions on University-Industry collaboration were optional (since not everyone would have engaged in such a collaboration) and were answered by nearly $65\%$ of the survey respondents. Table \ref{univindcoll} reveals that $12\%$ of the engagements were failure, $16\%$ successful and remaining partially successful. Table \ref{univindcoll} and Figure \ref{univind} shows respondents opinion on how to improve the collaboration between Industry and Academia. Respondents could select multiple options for this question. Figure \ref{univind} reveals that Industry sponsored PhD fellowships and encouraging student internships are enablers for improving the partnership. Difficulty in sharing Industrial data outside the company was selected as one of the major challenges and impediments encountered by the researchers in University-Industry collaboration.  
\begin{table*}
\centering
\caption{Data Sharing and Public Repositories}
\label{datasharing}
\begin{tabular}{|l|c|}
\hline
\multicolumn{2}{|p{16cm}|}{\textbf{[10] In MSR research have you ever made your dataset publicly available?}} \\ \hline
Never & 4.55\% \\
Rarely & 0\% \\
Sometimes & 59.09\% \\
Always & 36.36\% \\ \hline
\multicolumn{2}{|p{16cm}|}{\textbf{[11] Is your dataset freely available or do we need to request for permissions to use it?}} \\ \hline
Freely, in the public domain & 100\% \\
With a dataset sharing agreement or license & 0\% \\
With a service fee for use (by industry) to help maintain the dataset & 0\% \\ \hline
\multicolumn{2}{|p{16cm}|}{\textbf{[12] Which platform did you use to share your dataset?}} \\ \hline
Home Page & 63.64\% \\
GitHub & 45.45\% \\
Bitbucket & 9.09\% \\ 
Submitted to existing repository (eg. PROMISE Repository) & 4.55\% \\
Others (please specify) & 13.64\% \\ \hline
\multicolumn{2}{|p{16cm}|}{\textbf{[13] Why do you want to share your dataset in MSR?}} \\ \hline
To contribute to the replication of experiments & 100\% \\
To allow meta-analysis (combining the findings from independent studies) & 45.45\% \\
To get more Empirical Software Engineering researchers involved in the MSR research & 68.18\% \\ 
Others (please specify) & 4.55\% \\ \hline
\multicolumn{2}{|p{16cm}|}{\textbf{[14] Should dataset sharing be a mandatory condition for publication in MSR conference?}} \\ \hline
No & 9.09\% \\
Neutral & 59.09\% \\
Yes & 31.82\% \\ \hline
\end{tabular}
\end{table*}
\begin{figure}[t]
\centering
\includegraphics[width=0.9\linewidth]{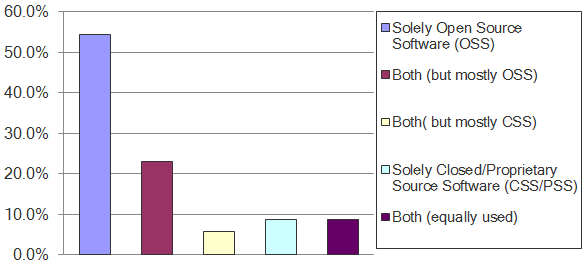}
\caption{Extent of OSS and CSS Dataset Usage in MSR Research Studies}
\label{osscss}
\end{figure}
\begin{table*}
\centering
\caption{OSS/CSS Dataset, Generalizability of Findings and Threats to External Validity}
\label{generalizability}
\begin{tabular}{|l|c|}
\hline
\multicolumn{2}{|p{16cm}|}{\textbf{[15] What type of dataset you have used in your MSR research study?}} \\ \hline
Solely Open Source Software (OSS) & 54.29\% \\
Both (but mostly OSS) & 22.86\% \\
Both( but mostly CSS) & 5.71\% \\
Solely Closed/Proprietary Source Software (CSS/PSS) & 8.57\% \\ 
Both (equally used) & 8.57\% \\ \hline
\multicolumn{2}{|p{16cm}|}{\textbf{[16] Do you believe that the results/findings on OSS dataset can be generalized to CSS dataset?}} \\ \hline
Never & 5.71\% \\
Rarely & 22.86\% \\
Sometimes & 71.43\% \\
Always & 0\% \\ \hline
\multicolumn{2}{|p{16cm}|}{\textbf{[17] Do you believe that within OSS dataset there is enough diversity for researchers to test for generalizability?}} \\ \hline
Never & 6.25\% \\
Rarely & 25.00\% \\
Sometimes & 62.50\% \\
Always & 6.25\% \\ \hline
\multicolumn{2}{|p{16cm}|}{\textbf{[18] Why do you believe that threats to external validity exists?}} \\ \hline
Lack of accessibility to CSS dataset & 75.00\% \\
Usage of few well-known and OSS dataset & 40.63\% \\
Others (please specify) & 6.25\% \\ \hline
\multicolumn{2}{|p{16cm}|}{\textbf{[19] How can we improve external validity concerns in MSR research?}} \\ \hline
Creating benchmark suite by the research community and sharing the analysis results on it & 62.50\% \\
Reviewers must discuss the validity concerns and evaluation criteria for paper selection & 31.25\% \\
Making the dataset publicly available & 81.25\% \\
Others (please specify) & 3.13\% \\ \hline
\multicolumn{2}{|p{16cm}|}{\textbf{[20] Can you suggest ways to increase the contribution of studies using CSS/PSS dataset in MSR research?}} \\ \hline
Industry - Academia collaboration should be more promoted & 65.71\% \\
Sharing of CSS/PSS dataset by anonymization & 54.29\% \\
Others (please specify) & 11.43\% \\ \hline
\end{tabular}
\end{table*}
\subsection{Data Sharing and SE Data Repositories}
Sharing of Software Engineering data and creation of SE data repositories for the purpose of conducting benchmarking, experimental and empirical studies is critical in a discipline like Empirical Software Engineering and Mining Software Repositories where the validity of the scientific results and conclusions is highly dependent on the underlying dataset used for experiments. There have been attempts for creation of public data repositories in Software Engineering field where researchers can upload real-world project data.  Cheikhi et al. present their analysis of two largest of the small number of software engineering repositories publicly available: the ISBSG Repository which contains datasets covering a considerable number of fields, and the PROMISE repository with its large number of different datasets \cite{cheikhi2013}. 

Cukic et al. mention that lack of publicly available SE datasets results in poorly validated models. Furthermore, they mention that dataset submission by organizations to public repositories is challenging due to the fact that public release of any data that could link a company with a negative image is a major deterrent for the company towards sharing their data \cite{cukic2005}. Fernandez-Diego et al. present an analysis of the potential and limitations of the International Software Benchmarking Standards Group (ISBSG) dataset. They study how and to what extent ISBSG has been used by researchers from the year 2000. Their analysis reveals that studies including dataset from ISBSG were published in 19 Journals and 40 Conferences \cite{fernandez2014}.

Table \ref{datasharing} reveals that $36.36\%$ of the respondents always make their dataset publicly available. It is interesting to note that all those who share their dataset make their dataset freely available and does not require any sharing fee or dataset sharing agreement or license. Table \ref{datasharing} and Figure \ref{condition} presents respondents opinion on whether dataset sharing should be a mandatory condition for publication in MSR conference. Our survey reveals that $31.82\%$ feel that it should be mandatory while $59.09\%$ are neutral. Questions $12$ and $13$ in Table \ref{datasharing} are questions in which the respondent can select multiple answers. We observe that sharing data on home-page rather than a public repository was the most common platform for making the data available. It is interesting to note that project web-hosting websites like GitHub and BitBucket are more widely used than well-known public repositories like PROMISE for sharing dataset. All most all of the respondents believe that they want to share their dataset with other researchers to encourage replication, meta-analysis and enable more Empirical Software Engineering researchers involved in Mining Software Repositories research.

\subsection{Replication and Threats to External Validity}
Robles et al. conduct a study of $171$ papers from six MSR conferences (year $2004$ to $2009$) that contained any experimental analysis of software projects for their potentiality of being replicated. Their findings show that MSR authors use in general publicly available data sources [such as data from OSS projects like Google Android and Chromium, Mozilla FireFox and Eclipse], mainly from free software repositories, but that the amount of publicly available processed datasets is very low. They also investigated the public availability of tools and scripts created by authors and show that for a majority of papers they were not able to find any tool, even for papers where the authors explicitly state that they have built one \cite{robles2010}. 

Shull et al. mention that reproducibility and replication in Empirical Software Engineering research is important and the two important goals of replication are to gain confidence in results of previous studies and also for understanding the scope of the results \cite{shull2008}. Barr et al. mention that Software engineering research will advance further and faster if the sharing of data and tools were easier and more widespread. They discuss pragmatic concerns such as the time and effort required and the risk of being scooped which hinder the realization of this idea of data sharing. They examine the costs and benefits of facilitating sharing in the field of Software Engineering in an effort to help the community understand what problems exist and find a solution \cite{barr2010}.

Table \ref{generalizability} shows the results of MSR authors on the topic of usage of OSS/CSS dataset, generalizability  of the experiments are conducted solely on OSS dataset. Answers to Question $15$ in Table \ref{generalizability} shows lack of empirical studies on closed or proprietary dataset (refer to Figure \ref{osscss}). Nearly $30\%$ of the respondents believe that the results and findings on OSS dataset can never or rarely be generalized to CSS dataset. Similarly, nearly $30\%$ of the respondents believe that there is not enough (rarely or never) diversity within OSS dataset for researchers to test for generalizability. 

Questions $18-20$ in Table \ref{generalizability} are questions in which the respondent can select multiple answers. $75\%$ of the respondents believe that lack of accessibility of CSS dataset as one of the major threats to external validity. We asked questions to MSR authors on how can we improve external validity concerns in MSR research and Can you suggest ways to increase the contribution of studies using CSS/PSS dataset in MSR research? Respondents mention that making the dataset publicly available and creating benchmark suite by the research community and sharing the analysis results on it can improve external validity concerns. Industry - Academia collaboration should be more promoted and sharing of CSS/PSS dataset by anonymization are ways to increase the contribution of studies using CSS/PSS dataset in MSR research. 

\section{Conclusion}
We conduct a survey of authors of past three years of MSR conference ($2012$, $2013$ and $2014$) to collect data on their views and suggestions to address the stated concerns. Nearly $75\%$ of the survey respondents were affiliated to a University whereas only $25\%$ respondents were from Industry. We received opinions from MS and PhD Scholars in University, Faculty Members, Researcher, Software Engineer and Manager in an Industry. Our survey reveals that Industry sponsored PhD fellowships and encouraging student internships are enablers for improving the partnership between Industry and Academia. Our findings shows that $31.82\%$ of the respondents feel that data sharing should be a mandatory condition for publication while $59.09\%$ are neutral.  All most all of the respondents believe that they want to share their dataset with other researchers to encourage replication, meta-analysis and enable more Empirical Software Engineering researchers involved in Mining Software Repositories research. Respondents mention that making the dataset publicly available and creating benchmark suite by the research community and sharing the analysis results on it can improve external validity concerns. Industry - Academia collaboration should be more promoted and sharing of CSS/PSS dataset by anonymization are ways to increase the contribution of studies using CSS/PSS dataset in MSR research.
\bibliographystyle{plain}
\bibliography{sen}
\end{document}